\documentclass{article}
\usepackage{amssymb}
\usepackage{amsfonts}
\usepackage{amsmath}

\input{tcilatex}

\begin{document}

\title{$\eta $-weak-pseudo-Hermiticity generators and exact solvability}
\author{Omar Mustafa$^{1}$ and S.Habib Mazharimousavi$^{2}$ \\
Department of Physics, Eastern Mediterranean University, \\
G Magusa, North Cyprus, Mersin 10,Turkey\\
$^{1}$E-mail: omar.mustafa@emu.edu.tr\\
$^{2}$E-mail: habib.mazhari@emu.edu.tr}
\maketitle

\begin{abstract}
Exact solvability of some non-Hermitian $\eta $-weak-pseudo-Hermitian
Hamiltonians is explored as a byproduct of $\eta $-weak-pseudo-Hermiticity
generators. A class of $V_{eff}\left( x\right) =V\left( x\right) +iW\left(
x\right) $ potentials is considered, where the imaginary part $W\left(
x\right) $ is used as an $\eta $-weak-pseudo-Hermiticity generator to obtain
exactly solvable $\eta $-weak-pseudo-Hermitian Hamiltonian models.

\end{abstract}

\section{Introduction}

Since the early years of quantum mechanics evolution the exact solvability
of quantum mechanical models have attracted much attention. Some exactly
solvable model have already become typical standard examples in quantum
mechanical textbooks. However, it was believed that the reality of the
spectra of the Hamiltonians, describing quantum mechanical models, is
necessarly attributed to their Hermiticity. It was the non-Hermitian $%
\mathcal{PT}$ -symmetric Hamiltonians' proposal by Bender and Boettcher [1]
that relaxed the Hermiticity condition as a necessity for the reality of the
spectrum [1-7]. Herein, $\mathcal{P}$ denotes the parity ($\mathcal{P}x%
\mathcal{P}=-x$) and the anti-linear operator $\mathcal{T}$ mimics the time
reflection ($\mathcal{T}i\mathcal{T}=-i$).

Recently, Mostafazadeh [8] has introduced a broader class of non-Hermitian
pseudo-Hermitian Hamiltonians (a generalization of $\mathcal{PT}$ -
symmetry, therefore). In these settings [8-19], a Hamiltonian $H$ is
pseudo-Hermitian if \thinspace it obeys the similarity transformation:%
\begin{equation}
\eta \,H\,\eta ^{-1}=H^{\dagger },
\end{equation}%
where $\eta $ is a Hermitian invertible linear operator and $(^{\dagger })$
denotes the adjoint. However, if $H$ \ is an $\eta $-pseudo-Hermitian with
respect to the nontrivial "metric"operator 
\begin{equation}
\eta =O^{\dagger }O,
\end{equation}%
for some linear invertible operator $O:\mathcal{H}{\small \rightarrow }%
\mathcal{H}$ ($\mathcal{H}$ is the Hilbert space), then its spectrum is real
and $H$ satisfies the intertwining relation%
\begin{equation}
\eta \,H=H^{\dagger }\,\eta ,
\end{equation}%
where $\left( \eta \,H\right) $ is also Hermitian. Moreover, one can even
relax $H$ to be $\eta $-weak-pseudo-Hermitian by not restricting $\eta \,$\
to be Hermitian (cf., e.g., Bagchi and Quesne [15]), and linear and/or
invertible (cf., e.g., Solombrino [15]). Nevertheless, Fityo [13] has
implicitly used $\eta $-weak-pseudo-Hermiticity\emph{\ }without
invertibility as a necessary condition on $O$ and hence on $\eta $.

Very recently we have\ followed Fityo's [13] $\eta $-weak-pseudo-Hermiticity
condition and introduced a class of spherically symmetric non-Hermitian
Hamiltonians and their $\eta $-weak-pseudo-Hermiticity generators\emph{\ }%
[13], where a generalization beyon the nodeless 1D state was proposed. The
same recipe was extended to deal with a class of $\eta $-\emph{w}%
eak-pseudo-Hermitian $d$-dimensional Hamiltonians for quantum particles
endowed with position-dependent masses [14].

In this work, we target exact solvability of some non-Hermitian $\eta $%
-weak-pseudo-Hermitian Hamiltonians as a byproduct of our $\eta $%
-weak-pseudo-Hermiticity geneerators discussed in [13]. In section 2, we
present our procedure for a class of non-Hermitian Hamiltonians with 1D
effective potentials of the form $V_{eff}\left( x\right) =V\left( x\right)
+iW\left( x\right) $, where $V\left( x\right) $ and $W\left( x\right) $ are
real valued functions. In the current settings, we use the imaginary part of
the effectve potential, $W\left( x\right) $, as an $\eta $%
-weak-pseudo-Hermiticity generator to obtain the real part $V\left( x\right) 
$ of some feasibly exactly-solvable effective potential $V_{eff}\left(
x\right) $. We present, in section 3, three $\eta $-weak-pseudo-Hermitian
examples: A $PT$-symmetric Scarf II potential [15], a $PT$-symmetric
periodic-type potential [16], and a non-$PT$-symmetric Morse potential [17].
We conclude in section 4.

\section{$\protect\eta $-weak-pseudo-Hermiticity generators}

In this section we consider a class of 1D non-Hermitian Hamiltonians (in $%
\hbar =2m=1$ units) of the form%
\begin{equation}
H=-\partial _{x}^{2}+V_{eff}\left( x\right) ;\text{ \ \ }V_{eff}\left(
x\right) =V\left( x\right) +iW\left( x\right) ,
\end{equation}%
Then $H$ has a real spectrum if and only if there is a linear operator $O:$ $%
H\longrightarrow H$ such that $H$ is $\eta $-weak-pseudo-Hermitian with the
linear operator%
\begin{equation}
O=\partial _{x}+F(x)+iG(x)\implies O^{\dagger }=-\partial _{x}+F(x)-iG(x)
\end{equation}%
where $F(x)$ and $G(x)$ are real-valued functions and $%
\mathbb{R}
\ni x\in \left( -\infty ,\infty \right) $. Equation (2), in turn, implies%
\begin{equation}
\eta =-\partial _{x}^{2}-2iG(x)\partial _{x}+F(x)^{2}+G(x)^{2}-F^{\prime
}(x)-iG^{\prime }(x),
\end{equation}%
where primes denote derivatives with respect to $x.$ Herein, the operators $O
$ and $O^{\dag }$ are two intertwining operators and the Hermitian operator $%
\eta $ leads to the intertwining relation (3) (cf, e.g.,[8,13]). Hence,
relation (3) would lead to%
\begin{equation}
W(x)=-2G^{\prime }(x),
\end{equation}%
\begin{equation}
F\left( x\right) ^{2}-F^{\prime }\left( x\right) =\frac{2G\left( x\right)
G^{\prime \prime }\left( x\right) -G^{\prime }\left( x\right) ^{2}+\alpha }{%
4G\left( x\right) ^{2}},
\end{equation}%
\begin{equation}
V(x)=\frac{2G\left( x\right) G^{\prime \prime }\left( x\right) -G^{\prime
}\left( x\right) ^{2}+\alpha }{4G\left( x\right) ^{2}}-G\left( x\right)
^{2}+\beta 
\end{equation}%
where $\alpha ,\beta \in 
\mathbb{R}
$ are integration constants.

Now we depart from our formal procedure in [13] and seek the real part, $%
V\left( x\right) ,$ of the effective potential, $V_{eff}\left( x\right) ,$
using the imaginary part, $W(x),$ as a generating function. Equation (7)
would therefore lead to 
\begin{equation}
G(x)=-\frac{1}{2}\int^{x}W(z)dz,
\end{equation}%
with an integration constant set equals zero for simplicity/convenience. In
a straightforward manner one can show, following (8) and (9) respectively,
that%
\begin{gather}
F(x)^{2}-F^{\prime }(x)=W^{\prime }(x)\left( 2\int^{x}W(z)dz\right)
^{-1}-\left( W(x)\left( 2\int^{x}W(z)dz\right) ^{-1}\right) ^{2}  \notag \\
+\,\alpha \left( \int^{x}W(z)dz\right) ^{-2}.
\end{gather}%
\begin{eqnarray}
V(x) &=&W^{\prime }(x)\left( 2\int^{x}W(z)dz\right) ^{-1}-\left( W(x)\left(
2\int^{x}W(z)dz\right) ^{-1}\right) ^{2}  \notag \\
&&+\,\alpha \left( \int^{x}W(z)dz\right) ^{-2}-\left( \frac{1}{2}%
\int^{x}W(z)dz\right) ^{2}+\beta 
\end{eqnarray}%
At this point, we should report that $\alpha ,\beta \in 
\mathbb{R}
$ can be used as adjustable real parameters that would serve for the exact
solvability of the $\eta $-weak-pseudo-Hermitian generators' productions of
the real part $V\left( x\right) $ of the effective potential $V_{eff}\left(
x\right) $.

\section{Illustrative examples}

In this section, we construct $\eta $-weak-pseudo-Herrmitian Hamiltonians
and exact solvability\ of some non-Hermitian Hamiltonians using our $\eta $%
-weak-pseudo-Herrmiticity generator$\ W(x)$ proposed above.

\subsection{An $\protect\eta $\emph{-}weak-pseudo-Herrmitian $PT$-symmetric
Scarf II Hamiltonian model }

An $\eta $-weak-pseudo-Herrmitian$\ $generator of the form%
\begin{equation}
W(x)=\frac{-A\sinh (x)}{\cosh ^{2}(x)}
\end{equation}%
would lead, using (12), to%
\begin{equation}
V(x)=-\frac{3+A^{2}}{4\cosh ^{2}(x)}.
\end{equation}%
Hence, the corresponding $\eta $\emph{-}weak-pseudo-Herrmitian Hamiltonian,
with $\alpha =0$ and $\beta =-1/4$, is given by%
\begin{equation}
H=-\partial _{x}^{2}-\frac{3+A^{2}}{4\cosh ^{2}(x)}-i\frac{A\sinh (x)}{\cosh
^{2}(x)}.
\end{equation}%
This Hamiltonian model represents an $\eta $-weak-pseudo-Herrmitian $PT$%
-symmetric Scarf II model which is exactly solvable (cf., e.g., Ahmet [15],
Khare [18], and Cooper et al [19]). The eigenvalues and eigenfunctions of
which are reported by Ahmet [15] as%
\begin{eqnarray}
\psi _{n}\left( x\right)  &=&C_{n}\left( \frac{1}{\cosh (x)}\right) ^{\left(
s+t-1/2\right) }  \notag \\
&&\times \exp \left[ \frac{i}{2}\left( t-s\right) \tanh ^{-1}\left( \sinh
\left( x\right) \right) \right] P_{n}^{-t-s}\left( i\sinh \left( x\right)
\right) 
\end{eqnarray}%
\begin{equation}
E_{n}=\left\{ 
\begin{tabular}{ll}
$-\left( n+\frac{1-A}{2}\right) ^{2}\,;\;n=0,1,\cdots <\frac{A-1}{2}%
;\smallskip \smallskip $ & for $A\geqslant 2,$ \\ 
$-\frac{1}{4}$ \ for $A<2.\smallskip $ & 
\end{tabular}%
\right. 
\end{equation}%
where 
\begin{equation}
s=\frac{1}{2}\left\vert A-2\right\vert \text{ \ \ and \ \ }t=\frac{1}{2}%
\left\vert A+2\right\vert 
\end{equation}

\subsection{An $\protect\eta $\emph{-}weak-pseudo-Herrmitian periodic-type $%
PT$-symmetric Hamiltonian model}

An $\eta $\emph{-}weak-pseudo-Herrmiticity generator of the form%
\begin{equation}
W(x)=\frac{4\sin (2x)}{3(\cos ^{2}(x)-\frac{4}{3})^{2}}
\end{equation}%
would result, with $\alpha =0$ \ and $\beta =1,$ in%
\begin{equation}
V(x)=\frac{1}{9}\frac{\left( -30\cos ^{2}(x)+24\right) }{\left( \cos ^{2}(x)-%
\frac{4}{3}\right) ^{2}}
\end{equation}%
This in turn yeilds an $\eta $-weak-pseudo-Herrmitian Hamiltonian%
\begin{equation}
H=-\partial _{x}^{2}-\frac{6}{\left( \cos (x)+2i\sin (x)\right) }.
\end{equation}%
The solution of which is reported by Samsonov [16], in the interval $x\in
\left( -\pi ,\pi \right) $ with the boundary conditions $\psi (-\pi )=\psi
(\pi )=0,$ as%
\begin{eqnarray}
\psi _{n}(x) &=&\left\{ \left[ \left( 16-n^{2}\right) \cos x-2i\left(
n^{2}-4\right) \sin x\right] \sin \left[ \frac{n}{2}\left( \pi +x\right) %
\right] \right.   \notag \\
&&\left. -6n\sin x\cos \left[ \frac{n}{2}\left( \pi +x\right) \right]
\right\} \times \left( \cos x+2i\sin x\right) ^{-1}
\end{eqnarray}%
\begin{equation}
E_{n}=n^{2}/4\text{ ; \ }n=1,3,4,5,...
\end{equation}%
with a missing state at $n=2$ (for more details the reader may refer to
Samsonov [16]).

\subsection{An $\protect\eta $-weak-pseudo-Herrmitian non-$PT$-symmetric
Morse Hamiltonian}

An $\eta $-weak-pseudo-Herrmiticity generator of the form 
\begin{equation}
W\left( x\right) =-\xi e^{-x}
\end{equation}%
would result, with $\alpha =0$ \ and $\beta =-1/4,$ in%
\begin{equation}
V(x)=-\frac{1}{4}\xi ^{2}e^{-2x}.
\end{equation}%
This is \emph{an }$\eta $\emph{-}weak-pseudo-Herrmitian non-$PT$-symmetric
Morse Hamiltonian\ model%
\begin{equation}
H=-\partial _{x}^{2}-\frac{1}{4}\xi ^{2}e^{-2x}-i\xi e^{-x},
\end{equation}%
considered by Ahmed [17], who has reported its eigenfunction and eigenvalue,
with $A=0,$ $B=\xi $ and $C=1/2$, as%
\begin{equation}
E_{0}=-1/4
\end{equation}%
$_{{}}$%
\begin{equation}
\psi _{n}\left( x\right) =z^{1/2}e^{-z/2}L_{0}^{1}(z)
\end{equation}%
where $z=2i\xi e^{-x}$.

\section{Conclusion}

In this work, we have introduced a byproduct of our $\eta $\emph{-}%
weak-pseudo-Herrmiticity generators discussed in [13]. We have shown that
the imaginary part, $W\left( x\right) $, of the effective potential, $%
V_{eff}\left( x\right) =V\left( x\right) +iW\left( x\right) $, can be used
as an $\eta $-weak-pseudo-Herrmiticity generator to come out with exactly
solvable Hamiltonian models. The utility of the current recipe is
demonstrated through a $PT$-symmetric Scarf II, a $PT$-symmetric
periodic-type, and a non-$PT$-symmetric Morse\ models.

Finally, we may report that although the choice of $W\left( x\right)
=W_{\circ }\in 
\mathbb{R}
$, where $W_{\circ }$ is constant, is feasible for our $\eta $%
-weak-pseudo-Herrmiticity generators, one should avoid such setting in $%
W\left( x\right) .$ This choice would result in 
\begin{equation*}
V_{eff}\left( x\right) =\frac{\alpha -W_{\circ }^{2}/4}{\left( W_{\circ
}x+C_{\circ }\right) ^{2}}-\frac{1}{4}\left( W_{\circ }x+C_{\circ }\right)
^{2}+iW_{\circ }+\beta 
\end{equation*}%
where $C_{\circ }\in 
\mathbb{R}
$ is an integration constant. It is obvious that such an effective potential
does not suport bound states (i.e., the spectrum discretness of the $\eta $%
\emph{-}pseudo-Herrmiticity required by Mostafazadeh's theorem in [8] is
violated). Of course, this is only valid for our settings of the $\eta $%
-weak-pseudo-Herrmiticity generators. That is, only for properly chosen $%
W\left( x\right) $ does the effective potential become exactly solvable and
there is no known systematic way of making such choices. It seems that this
is the only sacrifice we have to make for the sake of exactly solvable $\eta 
$-weak-pseudo-Herrmitian Hamiltonian models.

\vspace{0pt}

\vspace{0pt}

\end{document}